\documentclass[twocolumn,aps,amssymb,floatfix,superscriptaddress,preprintnumbers]{revtex4}
\usepackage[usenames]{color}

\usepackage{graphicx}
\usepackage{bm}
\usepackage{amsmath}
\usepackage{amssymb}
\usepackage{amsfonts}
\usepackage{float}
\usepackage{dsfont}  
\usepackage{slashed}  
\usepackage{booktabs}
\usepackage{multirow}
\usepackage{subfigure}
\usepackage{soul}
\usepackage[sort&compress]{natbib}

\usepackage{hyperref}
\usepackage{xr-hyper}
\makeatletter
\newcommand*{\addFileDependency}[1]{
  \typeout{(#1)}
  \@addtofilelist{#1}
  \IfFileExists{#1}{}{\typeout{No file #1.}}
}
\makeatother

\newcommand{\be}{\begin{equation}}  
\newcommand{\ee}{\end{equation}}  
\newcommand{\beq}{\begin{eqnarray}}  
\newcommand{\eeq}{\end{eqnarray}}

\newcommand{\tr}{\mathrm{Tr}}

\newcommand{\bea}{\begin{eqnarray}}
\newcommand{\eea}{\end{eqnarray}}

\newcommand{\MSb}{{\overline{\rm MS}}}

\begin{document}

\title{Flavor decomposition for the proton helicity parton distribution functions}


\author{Constantia Alexandrou}
\affiliation{
Department of Physics,
  University of Cyprus,
  P.O. Box 20537,
  1678 Nicosia,
  Cyprus}
\affiliation{
  Computation-based Science and Technology Research Center,
  The Cyprus Institute,
  20 Kavafi Street,
  Nicosia 2121,
  Cyprus}
\author{Martha Constantinou}
\affiliation{Department of Physics, Temple University, 1925 N. 12th Street, Philadelphia, PA 19122-1801, USA}
\author{Kyriakos Hadjiyiannakou}
\affiliation{
Department of Physics,
  University of Cyprus,
  P.O. Box 20537,
  1678 Nicosia,
  Cyprus}
  \affiliation{
  Computation-based Science and Technology Research Center,
  The Cyprus Institute,
  20 Kavafi Street,
  Nicosia 2121,
  Cyprus}
  \author{Karl Jansen}
  \affiliation{NIC, DESY, Platanenallee 6, D-15738 Zeuthen, Germany}
\author{Floriano Manigrasso}
\affiliation{
Department of Physics,
  University of Cyprus,
  P.O. Box 20537,
  1678 Nicosia,
  Cyprus}
\affiliation{Institut für Physik, 
Humboldt-Universität zu Berlin, Newtonstr.\ 15, 
12489 Berlin, Germany}
\affiliation{Dipartimento di Fisica, 
Università di Roma ``Tor Vergata'', 
Via della Ricerca Scientifica 1, 
00133 Rome, Italy\\
\vspace*{-0.25cm}
\centerline{\includegraphics[scale=0.19]{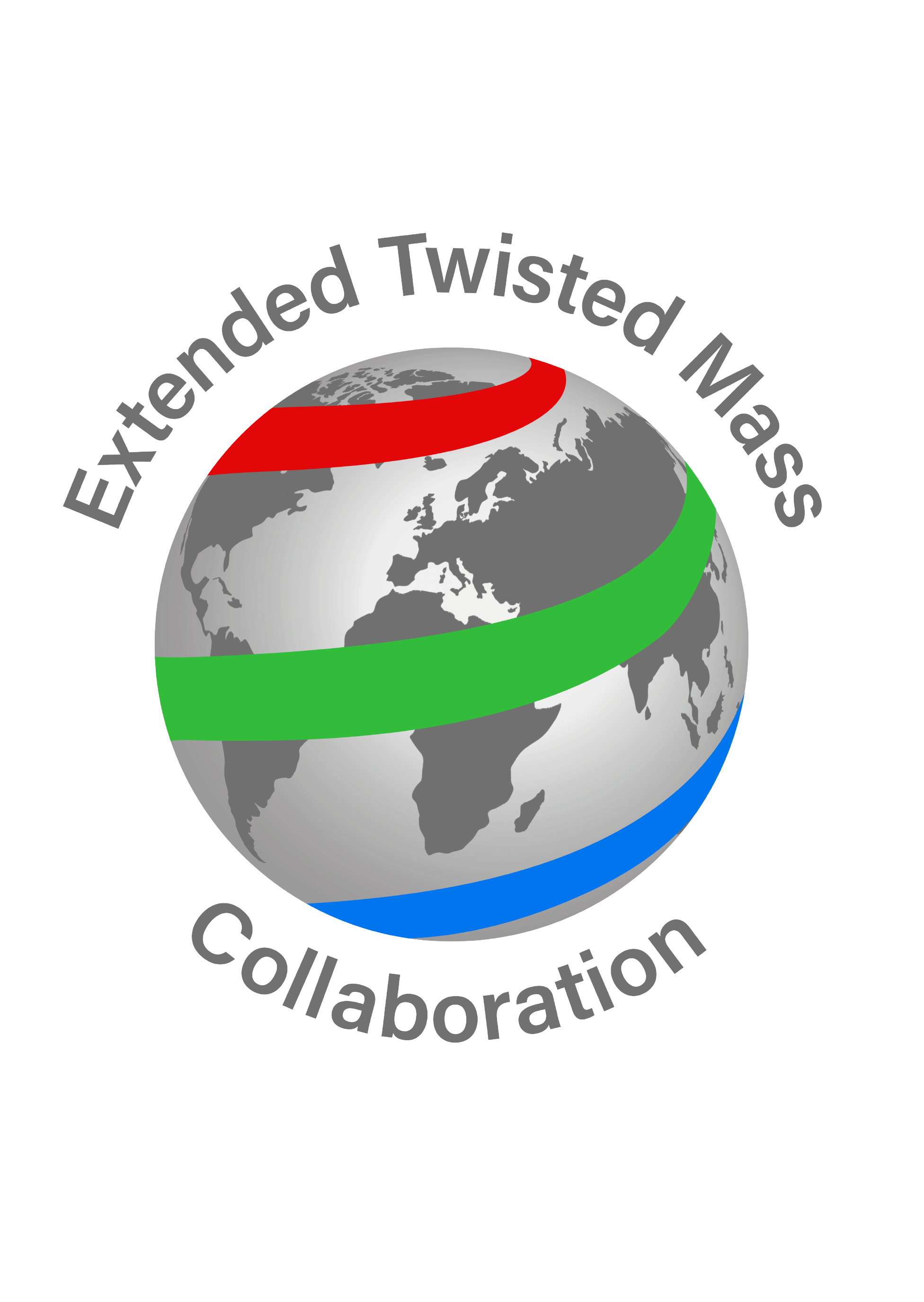}}
}

\begin{abstract}
\vspace*{-.7cm}
\noindent  We present, for the first time, an \textit{ab initio} calculation of the individual  up, down and strange quark helicity parton distribution functions for the proton.  
The calculation is performed within the twisted mass clover-improved fermion formulation of lattice QCD.  The analysis is performed using one ensemble of dynamical up, down, strange and charm  quarks with a pion mass of 260 MeV. The lattice matrix elements are  non-perturbatively renormalized and the final results are presented in the $\overline{ \rm MS}$ scheme at a scale of 2 GeV. We give results on the $\Delta u^+(x)$ and  $\Delta d^+(x)$, including disconnected quark loop contributions, as well as on the $\Delta s^+(x)$. For the latter we achieve  unprecedented precision compared to the phenomenological estimates.

\end{abstract}

\maketitle

\noindent
\textit{Introduction.}
\noindent
The theory of the strong interaction, Quantum Chromodynamics (QCD), successfully explains the structure of  hadrons and their interactions. The fundamental degrees of freedom in QCD are the quarks and gluons, collectively referred to as partons. Partons are responsible for the rich internal structure of hadrons. Most of the knowledge of the complex hadronic structure comes from parton distribution functions (PDFs), a set of number densities describing the non-perturbative QCD dynamics. Distribution functions are universal quantities and, therefore, can be accessed by a variety of high-energy scattering processes. The cross section of such processes can be factorized into a perturbative component calculable in perturbative QCD, and a non-perturbative part expressed in terms of the partonic densities. The generalized parton distributions (GPDs) and transverse momentum distribution functions (TMDs) complement PDFs, and are necessary for the 3-dimensional mapping of the hadrons.

At leading order within the parton model, the PDFs have a simple interpretation. The unpolarized PDFs are interpreted as the probability to find an unpolarized parton with a longitudinal momentum fraction $x$ within an unpolarized nucleon. The helicity PDFs can be interpreted as the difference between finding quarks with spins aligned and opposite to that of a longitudinally polarized nucleon. The colinear PDFs are completed with the transversity PDFs, which have the interpretation of finding quarks polarized in the same or opposite direction as a transversely polarized nucleon. PDFs play a central role in the on-going experimental program of major facilities, such as, BNL, CERN, DESY, Fermilab, JLab and SLAC (see, e.g., Refs.~\cite{Gao:2017yyd,Aidala:2012mv,Eskola:2016oht}). These experiments provide a wealth of measurements that are jointly analyzed within the framework of phenomenological fits. Based on the available experimental data, the most well-studied colinear distributions are the unpolarized, followed by the helicity with an order of magnitude less experimental data sets, namely a few hundred data sets~\cite{Ethier:2020way,Nocera:2014gqa}. The transversity PDFs are even less-known~\cite{Lin:2020rut}. The accessible kinematical region is more limited for the helicity and transversity PDFs as compared to the unpolarized PDFs, and therefore, the reconstruction of PDFs uses input from models. Such input introduces dependence on the functional forms employed. As a consequence, the extraction of the helicity and transversity PDFs are, to some extent, driven by the fit functions, due to the lack of experimental data (see, e.g., Ref.~\cite{Lin:2020rut} for a discussion). The dependence on the analysis procedure is evidence by the tension among some of the global analyses~\cite{Nocera:2014gqa,Ethier:2017zbq,deFlorian:2019zkl,Nocera:2017wep}. 

The focus of this work are the helicity PDFs, which are typically accessed experimentally in deep-inelastic scattering (DIS), semi-inclusive DIS, Drell-Yan, and proton-proton scattering processes. Currently, the global analyses use next-to-leading order (NLO) corrections in perturbative QCD (NNPDF$_{\rm POL}$1.1, DSSV14, JAM17)~\cite{Nocera:2014gqa,Ethier:2017zbq,deFlorian:2019zkl,Nocera:2017wep}. In these analyses, the up and down contributions, $\Delta u(x),\,\Delta d(x)$ are better constrained in the valence sector, with $\Delta u(x)$ being more precise. On the other hand, constraining $\Delta s(x)$ is not successful, as the kinematic regions of some of the data sets (e.g., the $W$-boson production data) are not sensitive to the strangeness~\cite{Nocera:2014gqa}. The situation somewhat improves with the inclusion of kaon production SIDIS data, but it is still unsatisfactory, and influenced by theoretical assumptions. In the recent work of the JAM Collaboration~\cite{Ethier:2017zbq} the authors used inclusive and semi-inclusive data, and  find, for both sets of data, the  strange polarization to be very small and consistent with zero. More details on the global analyses can be found in the recent reports~\cite{Lin:2017snn,Lin:2020rut}.

Based on the current status of phenomenological analyses, an extraction of the PDFs from \textit{first principles} is highly desirable. Here we present the first extraction of the up, down and strange helicity PDFs for the proton using lattice QCD, the only known \textit{ab initio} formulation of QCD.
We study both the valence and sea quark contributions that allow one to perform a controlled decomposition of the various distributions. To obtain the $x$-dependence of PDFs, we implement the quasi-PDF method~\cite{Ji:2013dva}. This approach is based on correlation functions that are calculable on a Euclidean lattice. The matrix elements are between include hadron state with  a finite momentum $\vec{P}=(0,0,P_3)$. A non-local operator with fermion fields separated by a distance $z$ connected by a Wilson line, is inserted between the proton states. Note that the Wilson line is in the same spatial direction as $\vec{P}$. Naturally, the matrix elements are defined in  coordinate space, with $z$ varying from zero up to half the spatial extent of the lattice. To extract physical quantities, a Fourier transform is applied on the matrix elements to obtain the so-called quasi-PDFs, which are defined in momentum space, $x$. For large values of $P_3$, the momentum boost in the quasi-PDFs can be interpreted as a Lorentz boost, recovering the light-cone PDF. The difference between quasi-PDFs and light-cone PDFs is ${\cal O}\left(\Lambda_{\rm QCD}^2/P_3^2,m_N^2/P_3^2\right)$ and is calculable in continuum perturbation theory within the Large Momentum Effective Theory (LaMET)~\cite{Ji:2014gla}. A successful research program on obtaining the PDFs using the quasi-PDFs method was developed since Ji's proposal, leading to theoretical and numerical advances~\cite{Chen:2016utp,Alexandrou:2016jqi,Briceno:2017cpo,Constantinou:2017sej,Alexandrou:2017huk,Ji:2017rah,Ji:2017oey,Ishikawa:2017faj,Green:2017xeu,Wang:2017qyg,Stewart:2017tvs,Izubuchi:2018srq,Alexandrou:2018pbm,Chen:2018fwa,Briceno:2018lfj,Spanoudes:2018zya,Alexandrou:2018eet,Liu:2018uuj,Radyushkin:2018nbf,Zhang:2018diq,Li:2018tpe,Alexandrou:2019lfo,Wang:2019tgg,Chen:2019lcm,Izubuchi:2019lyk,Cichy:2019ebf,Wang:2019msf,Son:2019ghf,Green:2020xco,Chai:2020nxw,Braun:2020ymy,Bhattacharya:2020cen,Bhattacharya:2020xlt,Bhattacharya:2020jfj,Chen:2020arf,Chen:2020iqi,Chen:2020ody,Ji:2020brr}. Recently, an exploratory study appeared on the strange and charm unpolarized PDFs~\cite{Zhang:2020dkn} using ensembles with pion mass 310 and 690 MeV. However, the work only presents matrix elements in coordinate space. Other methods on extracting the $x$-dependence of distribution functions have been discussed~\cite{Liu:1993cv,Detmold:2005gg,Braun:2007wv,Bali:2017gfr,Bali:2018spj,Detmold:2018kwu,Liang:2019frk,Ma:2014jla,Ma:2014jga,Radyushkin:2016hsy,Chambers:2017dov,Radyushkin:2017cyf,Orginos:2017kos,Ma:2017pxb,Radyushkin:2017lvu,Radyushkin:2018cvn,Zhang:2018ggy,Karpie:2018zaz,Sufian:2019bol,Joo:2019jct,Radyushkin:2019owq,Joo:2019bzr,Balitsky:2019krf,Radyushkin:2019mye,Sufian:2020vzb,Joo:2020spy,Bhat:2020ktg,Can:2020sxc,Alexandrou:2020zbe}. For an extensive review of the lattice calculations using the quasi-PDFs method, as well as other approaches to extract PDFs, see Refs.~\cite{Cichy:2018mum,Ji:2020ect}.

\vspace*{0.25cm}
\noindent
\textit{Lattice implementation.}
\noindent 
Based on the quasi-PDFs approach, the light-cone PDFs are obtained by the convolution of quasi-PDFs and the corresponding analytic expression for the matching kernel calculated in continuum perturbation theory. The quasi-PDFs are defined in momentum space as
\begin{equation}
\label{eq:quasi_pdf}
\widetilde{\Delta} q(x,\mu,P)= 2 P_3\, \int_{-\infty}^{+\infty}\frac{dz}{4\pi}\,
e^{-i x P_3 z}\,{\cal M}^R(z,P_3)\,,
\end{equation}
 and are Fourier transform of hadronic matrix elements
\begin{eqnarray}
{\cal M}^R(z,P_3,\mu) \equiv  Z(z,\mu) \, {\cal M}(z,P_3)\,, \qquad \qquad\\[2ex]
{\cal M}(z,P_3)\equiv \langle N(P)|\bar\psi\left(z\right)\gamma^3\gamma^5 W(0,z)\psi\left(0\right)|N(P)\rangle\,.
\label{eq:h}
\end{eqnarray}
These matrix elements are calculable on a Euclidean lattice. In Eq.~(\ref{eq:h}),
the proton initial and final states, $|N(P)\rangle$, carry the same momentum $P=(P_0, 0,0, P_3)$, as the PDFs are obtained in the forward kinematic limit. Here we focus on the helicity PDFs, $\Delta q\equiv g^q_1(x)$, and therefore we use the axial non-local operator, which contains a Wilson line $W(0,z)$ of length $z$ that guarantees gauge invariance. The bare matrix elements ${\cal M}(z,P_3)$ must be renormalized with an appropriate renormalization function, $Z(z,\mu)$, to remove divergences. We calculate $Z(z,\mu)$ using the RI$'$-type prescription proposed in Refs.~\cite{Constantinou:2017sej,Alexandrou:2017huk}
\begin{equation}
\label{renorm}
\frac{Z(z,\mu)}{12\, Z_q^{-1}(\mu)} 
{\rm Tr} \left[{\cal V}_{\gamma^3\gamma^5}(p,z) \left({\cal V}_{g_T}^{\rm Born}(p,z)\right)^{-1}\right] \Biggr|_{p^2{=}\mu_0^2}  \hspace*{-0.3cm}= 1\, ,
\end{equation}
which is applied at each value of $z$ separately. We refer the Reader to Ref.~\cite{Alexandrou:2019lfo} for notation. Due to the presence of the Wilson line in the operator, extracting the singlet renormalization functions is very challenging, as it involves a disconnected diagram. Here we use the non-singlet function indicated by $Z(z,\mu)$. We note that the difference between the singlet and non-singlet renormalization functions is expected to be small, as is the case  of the local axial-vector  operator~\cite{Alexandrou:2019brg}. This small difference has its origin to the fact that the difference between singlet and non-singlet arises to two loops in perturbation theory~\cite{Constantinou:2016ieh}. In addition to the logarithmic divergences and finite renormalization, the definition of Eq.~(\ref{renorm}) also removes the power-law divergence of the Wilson line. $Z(z,\mu)$ is obtained at an RI$'$ scale $\mu_0$. In our analysis, we convert to the $\overline{\rm MS}$ scheme at a scale $\mu=2$ GeV. An additional conversion factor is used to bring  $Z(z,\mu)$ in the modified $\overline{\rm MS}$-scheme~\cite{Alexandrou:2019lfo}. Therefore, the scale dependence appears in the renormalized matrix element ${\cal M}^R(z,P_3,\mu)$. While the matrix elements of local operators mix under renormalization~\cite{collins_1984}, the non-local operators under study do not mix in the renormalization process, as discussed in Refs.~\cite{Zhang:2018diq,Li:2018tpe,Wang:2019tgg}. This is because there is no additional non-local ultraviolet divergence in the quasi-PDF, an argument that holds to all orders in perturbation theory. However, the mixing occurs at the matching level and should be eliminated. To disentangle the singlet PDFs requires the matrix elements of the gluon PDFs, which is beyond the scope of this work. The nature of the mixing was also discussed earlier in Ref.~\cite{Green:2017xeu} using the auxiliary field approach.

The most widely-used method to obtain the quasi-PDFs is via the discretized Fourier transform of Eq.~(\ref{eq:quasi_pdf}). More recently, alternative  reconstruction techniques are being explored~\cite{Karpie:2018zaz,Karpie:2019eiq,Bhattacharya:2020cen,Alexandrou:2020tqq,Alexandrou:2020zbe}. In this work, we compare the standard Fourier transform, with the Bayes-Gauss-Fourier transform~\cite{Alexandrou:2020tqq}. We find agreement between the two approaches, indicating that the behavior of the lattice results at the large-$x$ region are not due to the discretization of the Fourier transform. We thus present  results using the discretized Fourier transform.

As can be seen in Eq.~(\ref{eq:quasi_pdf}), the quasi-PDFs depend on the nucleon momentum $P_3$, which should be finite but large. This dependence is expected to be removed by the matching kernel
\begin{equation}
\label{eq:matching}
\Delta q(x,\mu)=\int_{-\infty}^\infty 
\frac{d\xi}{|\xi|} \, C\left(\xi,\frac{\mu}{x P_3}\right)\, \widetilde{\Delta} q\left(\frac{x}{\xi},\mu,P_3\right)\,,  
\end{equation}
which is calculated to a given order in continuum perturbation theory. The matching kernel for the quasi-PDFs approach has been extensively studied (see, e.g., Refs.~\cite{Wang:2017qyg,Stewart:2017tvs,Izubuchi:2018srq,Bhattacharya:2020xlt,Bhattacharya:2020jfj,Chen:2020arf,Chen:2020iqi,Chen:2020ody}. In this work we employ the one-loop matching kernel in the modified $\overline{\rm MS}$-scheme, as defined in Ref.~\cite{Alexandrou:2019lfo}. Note that we choose the factorization scale to be the same as the renormalization scale $\mu$. The final step in extracting the light-cone PDFs is the application of the nucleon mass corrections, that have been calculated analytically in Ref.~\cite{Chen:2016utp}.

\vspace*{0.25cm}
\noindent
\textit{Numerical Methods.}
Obtaining ${\cal M}(z,P_3)$ is the most computationally demanding part of the calculation, as it contains a non-local operator, and must be calculated in the boosted frame. We perform the calculation including, for the first time, connected and quark-disconnected diagrams, for both the light and strange quark. In the light sector, we extract the isovector and isoscalar combinations, which are decomposed into the up and down  quark helicity PDFs. The calculation is performed using an ensemble of two light, a strange and a charm quark ($N_f=2+1+1$) within the twisted mass fermion formulation with clover term. The lattice spacing is $a=0.093$ fm and the lattice volume is 32$^3\times$64 ($L\approx3$ fm). The pion mass is about 260~MeV and $m_\pi L \approx 4$. 

The evaluation of the connected diagram uses the techniques outlined in Ref.~\cite{Alexandrou:2019lfo}, including the implementation of the momentum smearing method~\cite{Bali:2016lva}, and five stout smearing steps with parameter $\rho=0.15$, on the Wilson line entering the operator. Both smearing methods contribute to the reduction of the statistical noise. We refer to Ref.~\cite{Alexandrou:2019lfo} for the details.  We use a total number of measurements $N_{\rm meas}=392,\, 1552$ and 6320, for momenta $P_3=0.41,0.83$ and $1.24\,{\rm GeV}$, respectively. The source-sink separation is $t_s=0.94\,{\rm fm}$ for the lowest momentum and $t_s=1.13\,{\rm fm}$ for the other two.

The evaluation of the quark-disconnected diagrams involves the computation of disconnected quark loops that have to be combined with the nucleon two-point correlators. The disconnected quark loop with Wilson line reads
\begin{multline}\label{eq:loop}
    \mathcal{L}(t_{\rm ins},z) =\\
\sum\limits_{\vec{x}_{\rm ins}} \tr\left[D_q^{-1}(x_{\rm ins};x_{\rm ins}+z)\gamma^3\gamma^5 W(x_{\rm ins},x_{\rm ins}+z)\right],
\end{multline}
where $D_q^{-1}(x_{\rm ins};x_{\rm ins}+z)$ is the quark propagator, whose endpoints are connected by a Wilson line of length $z$. To reduce the stochastic noise coming from the low-modes~\cite{Abdel-Rehim:2016pjw}, we computed the first $N_{ev}=200$ eigen-pairs of the squared Dirac twisted-mass operator. From the eigen-pairs, the low-modes contribution to the all-to-all propagator can be exactly reconstructed and the high-modes contribution can then be evaluated with stochastic techniques. In particular, the stochastic evaluation of the disconnected loops is based on well-established techniques developed for local operators, such as hierarchical probing~\cite{Stathopoulos:2013aci}. The latter allows for reduction of the contamination of the off-diagonal terms in the evaluation of the trace of Eq.~\eqref{eq:loop}, up to a distance $2^k$. This is done using Hadamard vectors as basis vectors for the partitioning of the lattice. Here, the hierarchical probing algorithm has been implemented with $k=8$ in 4-dimensions, leading to 512 Hadamard vectors. 
In addition, for the stochastic evaluation of the disconnected loops in this work we make use of the \emph{one-end trick}~\cite{Abdel-Rehim:2013wlz,Alexandrou:2013wca} and fully dilute spin and color sub-spaces. We have recently employed successfully such methods in other studies of disconnected contributions. For more details see Refs.~\cite{Alexandrou:2020sml,Alexandrou:2019olr,Alexandrou:2019brg,Alexandrou:2018sjm}.  
\par
\vspace*{0.25cm}
\noindent
\textit{Results for the connected and disconnected contributions.}
For each value of the proton momentum, $P_3=0.41,0.83$ and $1.24\,{\rm GeV}$, we compute the two-point correlator for 200 source positions to reach a good  statistical accuracy. We also take all spatial orientations of $P_3$ and $W$, that is, $\pm x,\,\pm y,\, \pm z$. Moreover, both in the two-point and disconnected three-point functions we average over the forward and backward contributions. The total number of configurations analyzed is 330 for the two smallest momenta, and 480 for the largest one, bringing the total statistics to 66,000 and 96,000, respectively. Momentum smearing is applied for the two largest values $P_3=0.83,1.24\,{\rm GeV}$. The gauge links in the Wilson line entering the disconnected loop of Eq.~\eqref{eq:loop} undergo 10 iterations of stout smearing, with parameter $\rho=0.129$. 

To properly take into account the contamination of the excited states occurring at small source-sink separations $t_s$, we compute the disconnected three-point correlators at $t_s=0.75,0.84,0.94,1.03,1.13$~fm, and  perform a two-state fit analysis, following the procedure described in Ref.~\cite{Alexandrou:2020sml}. We find that the two-state fit gives results that are in agreement with those obtained form the plateau method analysis for $t_s=1.13$~fm. We will use the results from the plateau method for what follows. 
\begin{figure}[!htb]
    \centering
    \includegraphics[width=\linewidth]{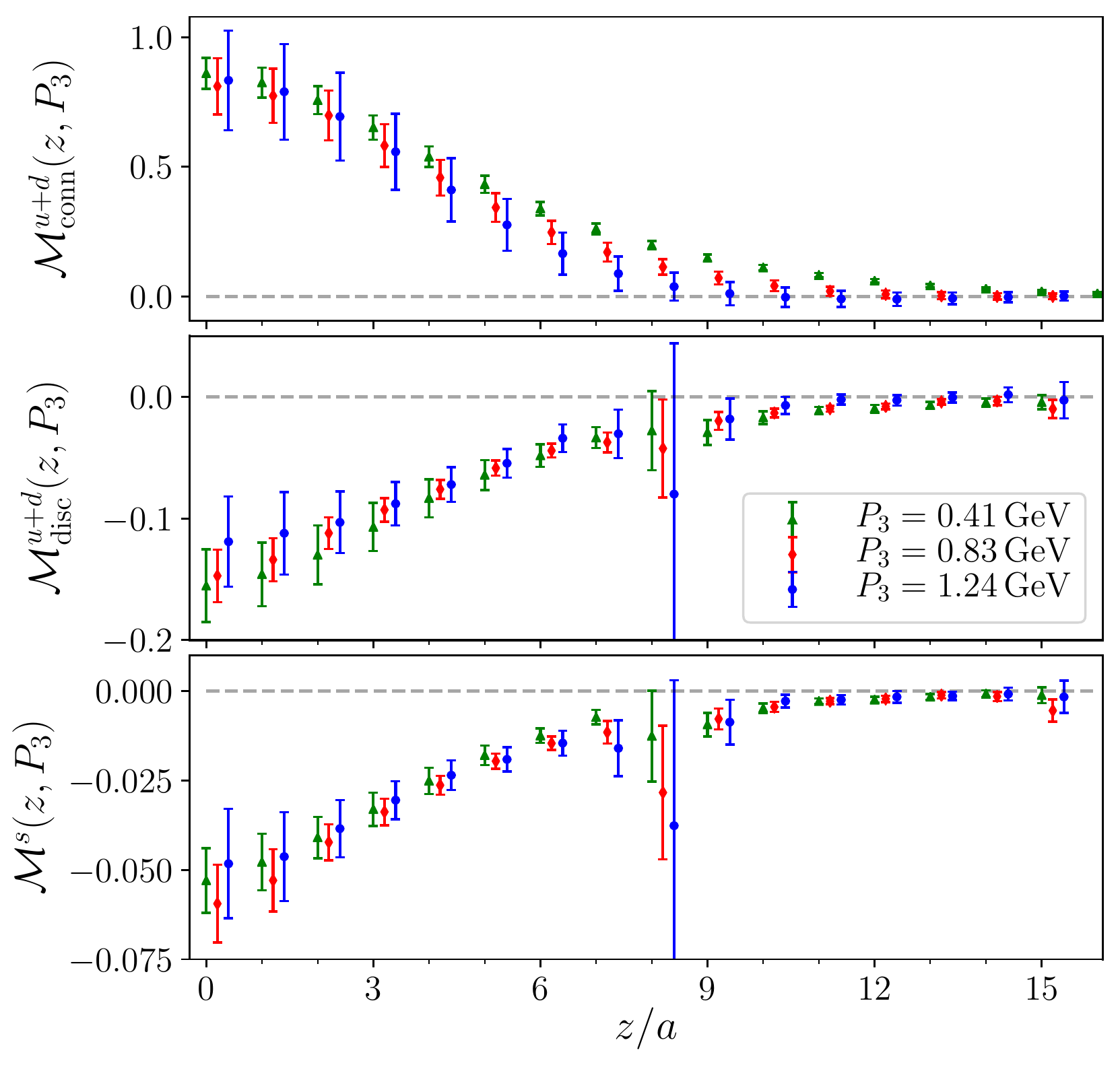}
    \caption{Real part of the bare matrix elements of the isoscalar $u+d$ connected (upper panel) and disconnected (middle panel) contributions and the strange quark (lower panel). The data for the disconnected matrix elements are obtained with the plateau fit performed at $t_s=1.13$~fm.}
    \label{fig:matrix_elements_disc}
\end{figure}
In Fig.~\ref{fig:matrix_elements_disc} we show the real part of the bare matrix elements using $t_s=1.13$~fm for the disconnected contributions. The disconnected part of  the isoscalar combination $u+d$,  is smaller than the connected isoscalar contribution as expected. The real strange matrix element is about  half as compared to the disconnected $u+d$. However, in both $u+d$ and strange  we clearly obtain a non-zero signal with the statistical uncertainties under control. The imaginary part of the bare disconnected matrix element is  compatible with zero at each $z$, and is not shown. The matrix elements smoothly decay to zero and for $z/a>8$ become compatible with zero. We note that the increase in the error for $z=8a$ in the disconnected part of the matrix element, i due to using hierarchical  probing with length $2^k$ and $k=3$. This is verified by repeating the evaluation of the disconnected diagrams with $k=2$, and  confirm that the same behavior occurs at $z=4a$ and its multiples, reflecting the limitation of the hierarchical probing technique when dealing with large lengths of the Wilson line. In taking  the Fourier transform in Eq. \eqref{eq:quasi_pdf}, we choose the cutoff $z_{max}$ such that the renormalized matrix element is compatible with zero. Since for  the isoscalar and isovector matrix elements this occurs  at different values of the Wilson line length $z$, we use different cutoffs $z_{max}$ for the two quantities. In particular, for the isoscalar case (the sum of connected and disconnected contributions) at $P_3=1.24$ GeV, we use $z_{max}=7a$, which is below the hierarchical probing length of $8a$. While, for the isovector case, the matrix element is compatible with zero at $z_{max}/a=9$.

Two additional important issues need to be addressed in order to extract the PDFs, namely the dependence of the results on the momentum boost and the accuracy of the discrete  Fourier transform. We examine these issues b considering the  $x \Delta d^+(x)\equiv x \left(\Delta d + \Delta\bar{d}\right)$ distribution, since the behavior is similar for the other two. 
  To extract the $\Delta d^+(x)$ distribution we apply renormalization and matching procedures  separately on the isovector, isoscalar and strange quasi-PDFs. As mentioned above, we neglect the mixing with the gluon PDFs at the matching level.

\begin{figure}[!htb]
    \centering
    \includegraphics[width=\linewidth]{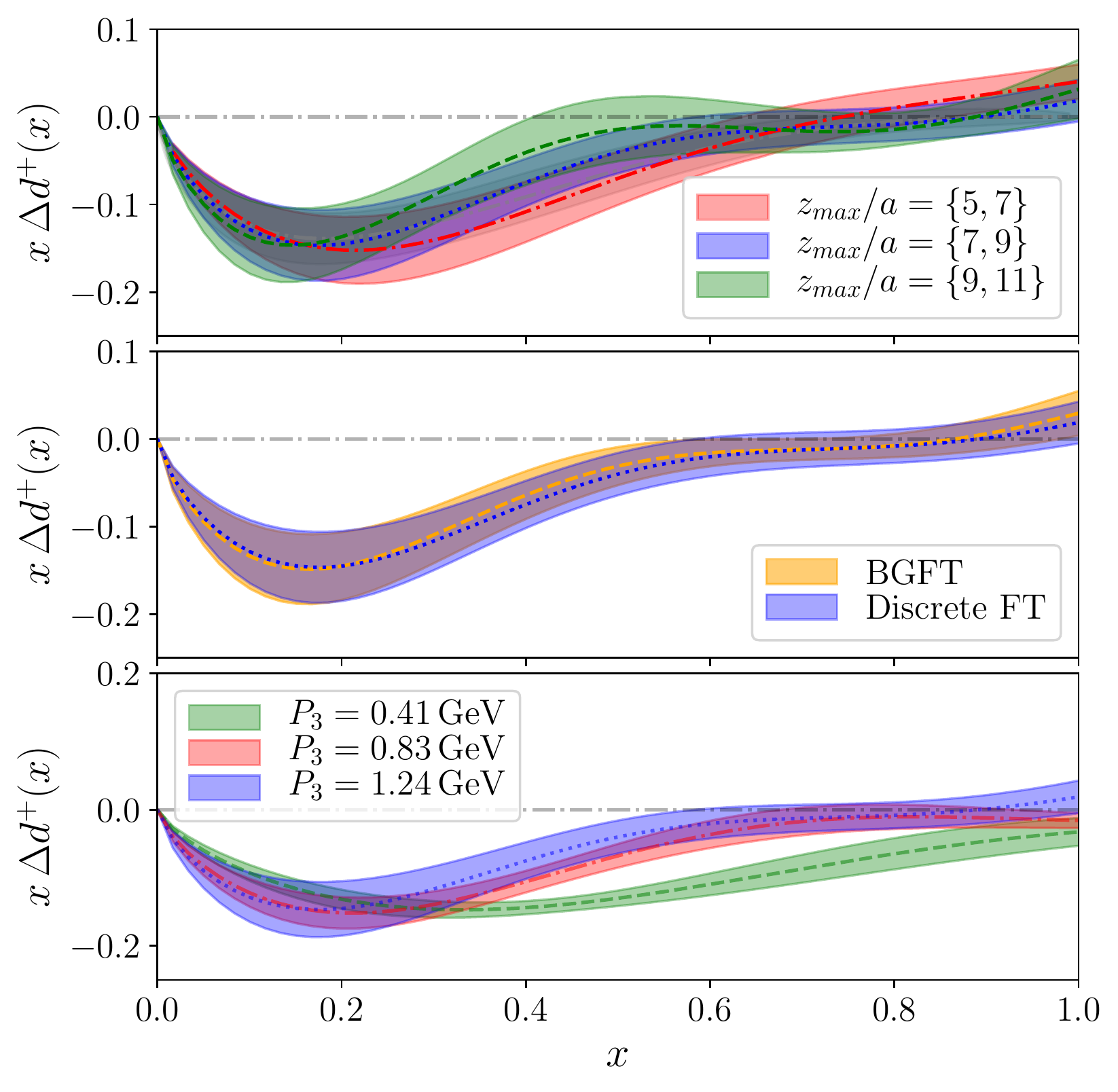}
    \caption{Dependence on the cutoff $z_{max}$ of the $\Delta d^+$ distribution (upper panel). The first(second) number reported between curly brackets indicates the value of $z_{max}$ adopted with the isoscalar(isovector) matrix element. Comparison between the $\Delta d^+$ distribution obtained with discrete Fourier Transform and with the BGFT technique \cite{Alexandrou:2020tqq} (middle panel). Momentum dependence of the distribution $\Delta d^+$  (bottom panel).}
    \label{fig:momentum_dependence_pdf}
\end{figure}

In Fig. \ref{fig:momentum_dependence_pdf} we show the momentum dependence of  $x \Delta d^+(x)$.  We observe that, while  when increasing the momentum from $0.41\,{\rm GeV}$ to $0.83\,{\rm GeV}$ there is a discrepancy in particular for large values of $x$, when we further increase the momentum to $P_3=1.24\,{\rm GeV}$, the results become compatible. This suggests that convergence has been reached within the limits of our current precision. In Fig. \ref{fig:momentum_dependence_pdf} we also show the dependence of the $ x\Delta d^+(x)$ distribution on the cutoff $z_{max}$ adopted in the computation of the isoscalar and isovector quasi-PDFs. Despite the fact that  when increasing $z_{max}$, the resulting distribution tends to show more pronounced oscillations, the results for differrent $z_{max}$ all agree within uncertainties. In order to estimate the extent of the systematic effect due to the discretization and truncation of the Fourier Transform (FT), we employ the Bayes-Gauss-Fourier Transform (BGFT) \cite{Alexandrou:2020tqq}. As can be seen in Fig. \ref{fig:momentum_dependence_pdf}, the distribution obtained with the BGFT technique is compatible with the standard reconstruction based on the discrete FT.

\begin{figure}[!htb]
    \centering
    \includegraphics[width=\linewidth]{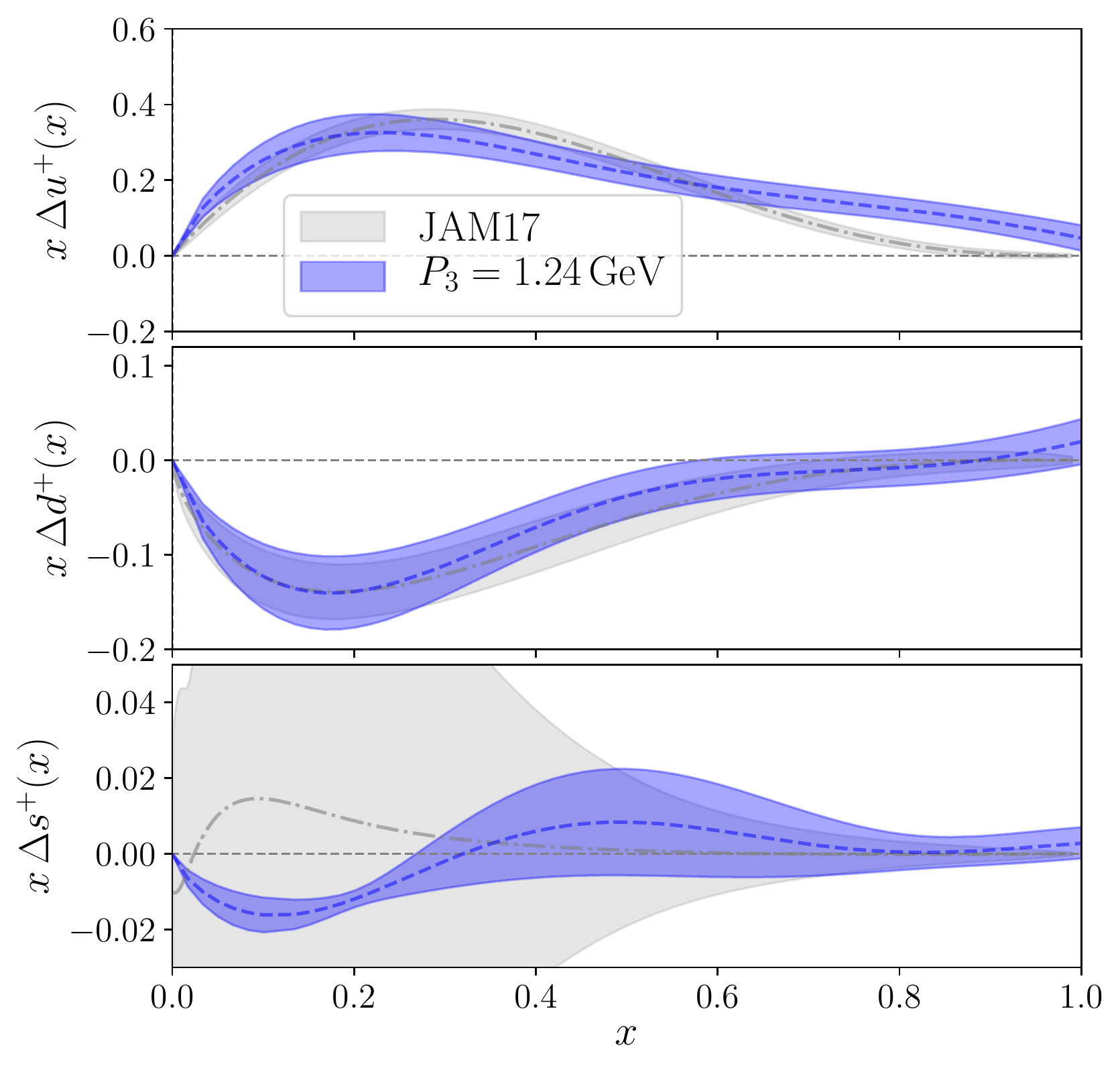}
    \caption{Comparison of lattice data on the up (upper), down (middle), and strange (bottom) quark helicity PDFs (blue) in the $\MSb$ scheme at $2\,{\rm GeV}$ with the JAM17 phenomenological data~\cite{Ethier:2017zbq} (gray).}
    \label{fig:flavor_decomposition}
\end{figure}

\vspace*{0.25cm}
\noindent
\textit{Flavor decomposition and comparison with phenomenology.} 
The aim of this work is to obtain the flavor decomposition of the up, down and strange quark distributions, by combining the total isoscalar and isovector contributions at each $P_3$ value. In Fig.~\ref{fig:flavor_decomposition} we show our final results at $P_3=1.24$ GeV for $|x| \Delta q^+(x)\equiv |x| \left(\Delta q + \Delta\bar{q}\right)$, for $q=u,\,d,\,s$, and compare with the JAM17 data~\cite{Ethier:2017zbq}. 
We find that   $x\Delta d^+(x)$ and $x\Delta s^+(x)$ nicely  decay to zero at $x=1$. While $x\Delta u^+(x)$ is also zero at $x=1$ and in agreement with the JAM17 results for $x\lesssim 0.6$, it decays slower than  the JAM17-determined distribution. On the other hand, we find a remarkable agreement for $x\Delta d^+(x)$ for the whole $x$ region. The lattice determination of the strange distribution $x\Delta s^+(x)$ is much more precise as compared to the one determined from the global analysis. Although small is non-zero for small values of $x$. This result provides a valuable input for phenomenological studies.

\vspace*{0.25cm}
\noindent
\textit{Conclusions.}
\noindent
Results for   the up, down and strange quark helicity PDFs of the proton, within lattice QCD are presented for the first time using the quasi-PDFs approach. We compute  matrix elements with nucleon states boosted to maximum momentum  $P_3=1.24$ GeV. We verify that the ground state matrix elements are well-determined by using one- and two-state fits, confirming that $t_s=1.13$ fm is sufficiently large to suppress excited state contributions  at this level of precision. The matrix elements are renormalized non-perturbatively, and matched to the light-cone PDFs using one-loop perturbation theory. For the flavor decomposition of the light quark PDFs we take into account, for the first time, both connected and disconnected diagrams and compute the totally disconnected strange PDF.   The final results on $|x| \Delta q^+$ are shown in Fig.~\ref{fig:flavor_decomposition}, and are compared with the global fits of the JAM Collaboration. We find a remarkable agreement for the case of $\Delta d^+$ for all values of $x$ and for case of  $\Delta u^+$ for $x<0.6$.  We also obtain $\Delta s^+$ much more precise that the phenomenological determination and show that is clearly non-zero for small values of $x$.  This work paves the way for a determination of these helicity PDFs using ensembles simulated with pion mass, which we plan to do in the near future. 

In the near future, a number of sources of systematic uncertainties will be explored, using the particular ensemble, along the lines of the analysis of Ref.~\cite{Alexandrou:2019lfo}. Other effects is the implementation of the mixing matching matrix between quark and gluon PDFs, that requires knowledge of the gluon matrix elements of non-local operators. Systematic uncertainties requiring more than one ensemble include discretization effects, volume effects, and pion mass dependence.
We plan to assess a proper determination of all sources of systematic uncertainties for the individual flavor PDFs in the future. Once systematic uncertainties are addressed and quantified, lattice results can provide useful input in the global fits for the strange PDFs, as well as the individual light-quark PDFs. This calculation is a first step towards achieving this goal.

\vspace*{0.5cm}
\begin{acknowledgements}

We would like to thank all members of ETMC for their constant and pleasant collaboration. We also thank N. Sato for providing the global fits data and J. Green for his comments. Finally, our thanks go to A. Scapellato for providing us the data at 0.43 GeV for the connected matrix elements. 
M.C. acknowledges financial support by the U.S. Department of Energy Early Career Award under Grant No.\ DE-SC0020405. K.H. is supported by the Cyprus Research and Innovation Foundation under grant POST-DOC/0718/0100. F.M. is supported  by the European Joint Doctorate program STIMULATE of the European Union’s Horizon 2020 research and innovation programme under grant agreement No 765048. 
This research includes calculations carried out on the HPC resources of Temple University, supported in part by the National Science Foundation through major research instrumentation grant number 1625061 and by the US Army Research Laboratory under contract number W911NF-16-2-0189. Computations for this work were carried out in part on facilities of the USQCD Collaboration, which are funded by the Office of Science of the U.S. Department of Energy. This research used resources of the Oak Ridge Leadership Computing Facility, which is a DOE Office of Science User Facility supported under Contract DE-AC05-00OR22725. The gauge configurations have been generated by the Extended Twisted Mass Collaboration on the KNL (A2) Partition of Marconi at CINECA, through the Prace project Pra13\_3304 "SIMPHYS". 
\end{acknowledgements}

\bibliography{references}

\end{document}